\documentclass[aps,prd,twocolumns,eqsecnum,preprintnumbers,showpacs,amsmath,amssymb]
{revtex4}

\usepackage{graphicx}
\usepackage{dcolumn}
\usepackage{bm}

\newcommand{\dd}{{\textrm d}}
\newcommand{\GeV}{{\textrm{GeV}}}
\newcommand{\MeV}{{\textrm{MeV}}}

\usepackage{floatrow}
\begin{document}

\title{THE HAGEDORN-TYPE STRUCTURE OF THE NON-PERTURBATIVE

GLUON PRESSURE WITHIN THE MASS GAP APPROACH TO QCD}

\vspace{3mm}

\author{V. Gogokhia$^1$, \ A. Shurgaia$^{2,3}$ \ M. Vas\'uth$^1$}
\email[]{gogohia.vahtang@wigner.mta.hu}
\email[]{avsh@rmi.ge}
\email[]{vasuth.matyas@wigner.mta.hu}

\vspace{3mm}

\affiliation{$^1$WIGNER RCP, RMI, Depart. Theor. Phys., Budapest 1121,
P.O.B. 49, H-1525, Hungary}

\affiliation{$^2$A. Razmadze Mathematical Inst. of I. Javakhishvili Tbilisi State University,
Depart. Theor. Phys., Tamarashvili str. 6, 0177 Tbilisi, Georgia}

\affiliation{$^3$European School, IB Word School, 34g A. Kazbegi Av., 0177 Tbilisi, Georgia}

\date{\today}
\begin{abstract}
We have shown in detail that the low-temperature expansion for the non-perturbative gluon pressure
has the Hagedorn-type structure. Its exponential spectrum of all the effective gluonic
excitations are expressed in terms of the mass gap.
It is this which is responsible for the large-scale dynamical structure
of the QCD ground state. The non-perturbative gluon pressure properly scaled has a maximum at some characteristic temperature
$T=T_c = 266.5 \ \MeV$, separating the low- and high temperature regions.
It is exponentially suppressed in the $T \rightarrow 0$ limit. In the $T \rightarrow T_c$ limit it demonstrates
an exponential rise in the number of dynamical degrees of freedom. Its exponential increase behavior with temperature is valid only
up to $T_c$. This makes it possible to identify $T_c$ with the
Hagedorn-type transition temperature $T_h$, i.e., to put $T_h=T_c$ within the mass gap approach to QCD at finite temperature.
The non-perturbative gluon pressure has a complicated dependence on the mass gap and temperature
near $T_c$ and up to approximately $(4-5)T_c$. In the limit of very high temperatures $T \rightarrow \infty$ its polynomial character is confirmed,
containing the terms proportional to $T^2$ and $T$, multiplied by the corresponding powers of the mass gap.
\end{abstract}



\maketitle

\section{Introduction}

The properties of Quantum Chromodynamics (QCD) at finite temperature and density are subject to the intense
investigations by lattice and analytic methods \cite{1,2,3,4} (and references therein).
The effective potential approach for composite operators \cite{5}
turned out to be effective and perspective analytical tool for the
generalization of QCD to non-zero temperature and density. In the
absence of external sources it is nothing but the vacuum energy
density (VED), i.e., the pressure apart from the sign. This approach is nonperturbative (NP) from the very
beginning, since it deals with the expansion of the corresponding
skeleton vacuum loop diagrams in powers of the Planck constant, and thus allows one to calculate the
VED from first principles. In accordance with this program we have
extended \cite{6} to non-zero temperature $T$ in \cite{7}.
This made it possible to introduce the correctly defined temperature-dependent bag
constant (bag pressure) as a function of the mass gap $\Delta$. It is this which is
responsible for the large-scale dynamical structure of the QCD ground state \cite{8} and
coincides with the Jaffe-Witten (JW) mass gap \cite{9} by properties.
The confining dynamics in the gluon matter (GM) is therefore nontrivially taken
into account directly through the mass gap and via the
temperature-dependent bag constant itself (denoted as $B_{YM}(T)$, see below) but other NP effects due to the mass gap are
also present (denoted as $P_{YM}(T)$, see below). Being NP, the effective approach for composite operators,
nevertheless, makes it possible to incorporate the thermal
perturbation theory (PT) expansion in a self-consistent way.
In our auxiliary work \cite{10} we have formulated and developed the analytic thermal PT which allows one to
calculate the PT contributions in terms of the convergent series in integer powers of a small
$\alpha_s$. We have also explicitly derived the first PT correction of the $\alpha_s \Delta^2 T^2$-order
to the purely NP part of the gluon pressure. We call the sum of all the calculated terms as the NP gluon pressure,
denoting it as $P_g(T)$ in what follows.

Here we are investigating a system at non-zero temperature, which consists of
$SU(3)$ purely Yang-Mills (YM) gauge fields without quark degrees of freedom (i.e., at zero density).
The primary aims of this article are to explicitly derive and briefly discuss the Hagedorn-type \cite{11} (and references therein) dynamical structure
of the above-mentioned NP gluon pressure below some characteristic temperature $T_c = 266.5 \ \MeV$, see fig. 1. These important issues
have been missed in our previous investigations \cite{7,8}.

\section{The gluon pressure at non-zero temperature}

For the readers convenience in order to have a general picture at hand, we begin with short sects. II, III, and IV,
in which we briefly describe our results obtained earlier in \cite{7,8}, especially taking into account that the book \cite{8}
is not freely available. In the imaginary-time formalism \cite{12,13,14}, all the
four-dimensional integrals can be easily generalized to non-zero temperature $T$ according to the prescription

\begin{equation}
\int {\dd q_0 \over (2\pi)} \rightarrow T \sum_{n=- \infty}^{+
\infty}, \quad \ q^2 = {\bf q}^2 + q^2_0 = {\bf q}^2 +
\omega^2_n = \omega^2 + \omega^2_n, \ \omega_n = 2n \pi T,
\end{equation}
i.e., each integral over $q_0$ of the loop momentum is to be
replaced by the sum over the Matsubara frequencies labeled by
$n$, which obviously assumes the replacement $q_0 \rightarrow
\omega_n= 2n \pi T$ for bosons (gluons). Let us also remind that in all our publications
as  well as in this paper the signature is always Euclidean in order to avoid non-physical singularities at light-cone
from the very beginning.

Introducing the temperature dependence into the NP gluon pressure \cite{7,8,10}, we obtain

\begin{equation}
P_g(T) = P^g_{NP}(T) + P_M(T) = B_{YM}(T) + P_{YM}(T) + P_M(T),
\end{equation}
where the corresponding terms in frequency-momentum space are:

\begin{equation}
B_{YM}(T) = { 8 \over \pi^2} \int_0^{\omega_{eff}} \dd\omega \
\omega^2 \ T \sum_{n= - \infty}^{+ \infty} \left[ \ln \left( 1 + 3
\alpha^{INP}(\omega^2, \omega^2_n) \right) - {3 \over 4}
\alpha^{INP}(\omega^2, \omega^2_n) \right],
\end{equation}

\begin{equation}
P_{YM}(T) = - { 8 \over \pi^2} \int_0^{\infty} \dd\omega \ \omega^2
\ T \sum_{n= - \infty}^{+ \infty} \left[ \ln \left( 1 + {3 \over 4}
\alpha^{INP}(\omega^2, \omega^2_n) \right) - {3 \over 4}
\alpha^{INP}(\omega^2, \omega^2_n) \right],
\end{equation}

\begin{equation}
P_M(T) = - { 8 \over \pi^2} \int_{\Lambda_{YM}}^{\infty}
\dd\omega \ \omega^2 \ T \sum_{n= - \infty}^{+ \infty} \left[ \ln
\left( 1 + {3 \alpha^{PT}(\omega^2, \omega^2_n) \over 4 + 3
\alpha^{INP}(\omega^2, \omega^2_n)} \right) - {3 \over 4}
\alpha^{PT}(\omega^2, \omega^2_n) \right].
\end{equation}

In frequency-momentum
space the intrinsically nonperturbative (INP) and PT effective charges become

\begin{equation}
\alpha^{INP}(q^2) = { \Delta^2 \over q^2} =
\alpha^{INP}( \omega^2, \omega_n^2) = { \Delta^2 \over \omega^2 + \omega_n^2},
\end{equation}
and

\begin{equation}
\alpha^{PT}(q^2) =  { \alpha_s \over 1 + \alpha_s b_0 \ln ( q^2 / \Lambda^2_{YM})}  =
\alpha^{PT} (\omega^2, \omega_n^2) = { \alpha_s \over 1 + \alpha_s b_0 \ln ( \omega^2
+ \omega_n^2 / \Lambda^2_{YM})},
\end{equation}
respectively. The last term (2.5) is called mixed (M) since it depends on both effective charges.
It is also convenient to introduce the following standard notations:

\begin{equation}
T^{-1} = \beta, \quad \omega = \sqrt{{\bf q}^2},
\end{equation}
where, evidently, in all the expressions ${\bf q}^2$ is the square of
the three-dimensional loop momentum, in complete agreement with the relations (2.1), and $\omega_{eff}$ is a scale separating the low-
and high frequency-momentum regions.

In eq.~(2.6) $\Delta^2$ is the mass gap, mentioned above, which is responsible for the large-scale dynamical
structure of the QCD vacuum, and thus determines the scale of its NP dynamics. We have shown that confining
effective charge (2.6), and hence its $\beta$-function, is a result
of the summation of the skeleton (i.e., NP) loop diagrams, contributing to the full gluon self-energy in the $q^2 \rightarrow 0$ limit (the strong coupling regime for the effective charge). This summation has been performed within the corresponding equation of motion. It has been done without violating the $SU(3)$ color gauge invariance of QCD \cite{8} (and references therein). In more detail the derivation of the bag constant as a function of the mass gap and its generalization to non-zero temperature has been completed in \cite{6} and \cite{7}, respectively.

The PT effective charge $\alpha^{PT}(q^2)$ (2.7) is the generalization to non-zero temperature of the
renormalization group equation solution, the so-called sum of the main PT logarithms \cite{8,15,16,17} (its analog as a function of the variable
$T/T_c$ (see below) can be found, for example in \cite{8,13,18}).
Here $\Lambda^2_{YM} = 0.09 \ \GeV^2$ \cite{19} is the asymptotic scale parameter for $SU(3)$ YM fields,
and $b_0=(11 / 4 \pi)$ for these fields, while the strong fine-structure constant is
$\alpha_s \equiv \alpha_s(m_Z) = 0.1185(6)$ \cite{20}. In eq.~(2.7) $q^2$ cannot go below $\Lambda^2_{YM}$, i.e.,
$\Lambda^2_{YM} \leq q^2 \leq \infty$, which has already been
symbolically shown in eq.~(2.5). It is worth reminding that the separation between effective charges (2.6) and (2.7),
is not only exact but it is unique as well \cite{6,7,8}.

The purely NP pressure $P^g_{NP}(T) = B_{YM}(T)+ P_{YM}(T)$ and the mixed
pressure $P_M(T)$, and hence the NP gluon pressure $P_g(T)$ (2.2) itself,
are normalized to zero when the interaction is formally switched
off, i.e., letting $\alpha_s = \Delta^2=0$. This means that the
initial normalization condition of the free PT vacuum to zero also holds
at non-zero temperature. Concluding, it is worse emphasizing that the bag pressure (2.3), correctly defined as the
difference between the PT and NP VED \cite{6,7,8}, and thus it is free of all the types of the PT contributions ("contaminations").

\section{The purely NP gluon pressure $P^g_{NP}(T)$ }

One of the attractive features of the confining
effective charge (2.6) is that it allows an exact summation over the Matsubara
frequencies in the purely NP pressure $P^g_{NP}(T)$ given by the sum of the integrals (2.3) and (2.4). Collecting all
the analytical results obtained in \cite{7,8}, we can write

\begin{equation}
P^g_{NP}(T)= B_{YM}(T)+ P_{YM}(T) = {6 \over \pi^2} \Delta^2 P_1 (T) + {16 \over \pi^2} T N(T).
\end{equation}
Here $P_1(T)$ and $N(T)$ are

\begin{equation}
P_1(T) = \int_{\omega_{eff}}^{\infty} \dd \omega {\omega \over e^{\beta\omega} -1},
\end{equation}
and

\begin{equation}
N(T) = [P_2(T) + P_3(T) - P_4(T)],
\end{equation}
respectively, while

\begin{eqnarray}
P_2(T) &=& \int_{\omega_{eff}}^{\infty} \dd \omega \ \omega^2
\ln \left( 1- e^{-\beta\omega} \right), \nonumber\\
P_3(T)&=& \int_0^{\omega_{eff}} \dd \omega \ \omega^2 \ln
\left( 1 - e^{- \beta\omega'} \right), \nonumber\\
P_4(T) &=& \int_0^{\infty} \dd \omega \ \omega^2 \ln \left( 1 -
e^{- \beta \bar \omega} \right),
\end{eqnarray}
and $\omega'$ and $\bar \omega$ are given by the relations

\begin{equation}
\omega' = \sqrt{\omega^2 + 3 \Delta^2} = \sqrt{\omega^2 + m'^2_{eff}},
\quad \bar \omega = \sqrt{\omega^2 + {3 \over 4} \Delta^2} = \sqrt{\omega^2 + \bar m^2_{eff}}.
\end{equation}
It is worth reminding that in the purely NP gluon pressure (3.1)
the bag pressure $B_{YM}(T)$ (2.3) is responsible for the formation
of the massive gluonic excitations $\omega'$, while the YM part $P_{YM}(T)$ (2.4) is responsible for the formation of the
massive gluonic excitations $\bar \omega$.

The so-called gluon mean number \cite{12}, also known as Bose-Einstein distribution, is

\begin{equation}
N_g \equiv N_g(\beta, \omega) = {1 \over e^{\beta\omega} -1},
\end{equation}
where $\beta$ and $\omega$ are defined in eq.~(2.8). It
appears in the integrals (3.3)-(3.4) and describes the
distribution of the massless gluons in the medium. Replacing
$\omega$ by  $\bar \omega$ and $\omega'$ we can consider the
corresponding gluon mean numbers as describing the distribution and
correlation of the corresponding massive gluonic excitations in the medium, see
integrals $P_3(T)$ and $P_4(T)$ in eqs.~(3.4). They are of NP
dynamical origin, since their corresponding effective masses are due to the mass gap, namely $m'_{eff} = \sqrt{3} \Delta$
and $\bar m_{eff} = (\sqrt{3}/2) \Delta$, see (3.5).
All three different gluon mean numbers range
continuously from zero to infinity \cite{12}. We have the two
different massless excitations, propagating in accordance with the
integral (3.2) and the first of the integrals (3.4). However, they
are not free, since in the PT $\Delta^2=0$ limit they vanish (the
composition (3.3) becomes zero in this case).
So the purely NP pressure $P^g_{NP}(T)$ (3.1) describes the four different effective gluonic excitations.
The gluon mean numbers are closely related to the thermodynamic observables,
especially to the pressure. Its exponential
suppression in the $T \rightarrow 0$ limit and the polynomial
structure in the $T \rightarrow \infty$ limit are determined by the
corresponding asymptotics of the gluon mean numbers, see below.

Concluding, let us emphasize that the effective scale $\omega_{eff}$ is not an independent scale parameter.
Due to extremization of the mass gap-dependent effective potential,
from the stationary condition at zero temperature in \cite{6} and the scale-setting scheme at non-zero temperature in \cite{7} it follows that

\begin{equation}
\quad \omega_{eff} = 1.48 \Delta, \quad \Delta = 0.6756 \ \GeV.
\end{equation}
So it is expressed in terms of the initial fundamental and unique mass scale parameter in our approach - the mass gap $\Delta$
(for simplicity, its squared version $\Delta^2$ is conventionally called the mass gap as well throughout this paper).
The introduction of $\omega_{eff}$ is also convenient from the technical point of view in order to simplify our expressions, which otherwise would be rather cumbersome.

\section{Thermal PT }

One of our primary goals in \cite{10} was to develop the
analytic formalism for the numerical calculation of the mixed term (2.5). It made it possible to calculate the PT
contributions to the gluon pressure (2.2) in terms of the
convergent series in integer powers of a small $\alpha_s$. For this
goal, it is convenient to re-write the integral (2.5) as follows:

\begin{equation}
P_M(T) = - { 8 \over \pi^2} \int_{\Lambda_{YM}}^{\infty}
\dd\omega \ \omega^2 \ T \sum_{n= - \infty}^{+ \infty} \left[ \ln [
1 + x(\omega^2, \omega^2_n)] - {3 \over 4} \alpha^{PT}(\omega^2,
\omega^2_n) \right],
\end{equation}
where

\begin{equation}
x(\omega^2, \omega^2_n) = {3 \alpha^{PT}(\omega^2, \omega^2_n) \over 4 + 3
\alpha^{INP}(\omega^2, \omega^2_n)} = {3 \over 4} {( \omega^2 + \omega^2_n) \over M (\bar \omega^2, \omega_n^2)}
{ \alpha_s \over (1 +  \alpha_s \ln z_n)}
\end{equation}
with the help of the expressions (2.6) and (2.7), and where

\begin{equation}
M(\bar \omega^2, \omega^2_n) = \bar \omega^2 + \omega^2_n, \quad
\ln z_n \equiv \ln z (\omega^2,\omega^2_n) = b_0 \ln [(\omega^2 + \omega^2_n)/ \Lambda^2_{YM}],
\end{equation}
and $\bar \omega^2$ is given in eq.~(3.5). Let us also note that in these notations $\alpha^{PT}(\omega^2, \omega^2_n)$ shown in eq.~(2.7), becomes

\begin{equation}
\alpha^{PT}(\omega^2, \omega^2_n) \equiv \alpha (z_n) = { \alpha_s \over (1 +  \alpha_s \ln z_n)}.
\end{equation}

Collecting all the results obtained in \cite{8,10}, where it has been explicitly shown that variable $x(\omega^2, \omega^2_n)$
is always very small, we are able to present the mixed part of the gluon pressure (4.1) as a sum of the two terms, namely

\begin{equation}
P_M(T) = P^s_{NP}(T) + O(\alpha_s),
\end{equation}
where $O({\alpha_s})$ describes the terms of the PT origin, not explicitly considered here.
It does not make any sense indeed to discuss them unless the main PT contribution - the Stefan-Boltzmann (SB) pressure - is not taken into account
(see discussion in sects. VII and VIII). It is zeroth order term relatively to the $O(\alpha_s)$ terms in eq.(4.5). The PT corrections to the purely NP pressure denoted above as $P^s_{NP}(T)$ are given by the series as follows:

\begin{equation}
P^s_{NP}(T) =  \sum_{k=1}^{\infty} \alpha_s^k P_k (\Delta^2; T)
\end{equation}
with

\begin{equation}
P_k(\Delta^2; T) = { 9 \over 2 \pi^2} \Delta^2
\int_{\Lambda_{YM}}^{\infty} \dd\omega \ \omega^2 \ T \sum_{n= -
\infty}^{+ \infty} \left[ { 1 \over M(\bar \omega^2, \omega^2_n)}
(-1)^{k-1} \ln^{k-1} z_n \right].
\end{equation}

Here $P^s_{NP}(T)$ (4.6) describes the $\Delta^2$-dependent contribution, beginning with the
$\alpha_s$-order term. The whole expansion (4.6) is the correction in integer powers of $\alpha_s$ to the purely NP pressure $P^g_{NP}(T)$ (3.1),
which, in fact, is the zeroth order term.
It is worth emphasizing that all the series which arise from the initial term (4.1)
in integer powers of a small $x(\omega^2, \omega^2_n)$ (or, equivalently, $\alpha_s$ in eq.~(4.5))
are convergent \cite{8,10}. Thus, the approximation of the PT effective charge by the summation of the main
PT logarithms (2.7) is fully sufficient to calculate all the PT corrections to leading orders in powers of a small $\alpha_s$.
Concluding this section, let us note that in order to clarify and simplify notations of \cite{7,8,10} we change
notation $P_{NP}(T)$ there to $P^g_{NP}(T)$ here. Also $P_{PT}(T)$ to $P_M(T)$, $P_{PT}(\Delta^2;T)$ to $P^s_{NP}(T)$,
while retaining the same notation for $P_g(T)$.

\section{The NP gluon pressure $P_g(T)$}

Taking into account the above-mentioned remarks and eq.~(4.5), the gluon pressure (2.2) then becomes
the sum of the two terms (see just below). In the integral (4.7) for $k=1$ the summation over the Matsubara frequencies can be performed
analytically, i.e., exactly \cite{7,8,10}. So finally for the NP gluon pressure $P_g(T)$, one obtains

\begin{equation}
P_g(T) = P^g_{NP} (T) + P^s_{NP}(T),
\end{equation}
on account of eqs.~(3.1)-(3.4) and eq.~(4.5) with omitting of the $O({\alpha_s})$-term, as discussed above, while

\begin{equation}
P^s_{NP}(T) =  \alpha_s \times {9 \over 2 \pi^2} \Delta^2
\int_{\Lambda_{YM}}^{\infty} \dd\omega \ \omega^2 \
{ 1 \over \bar \omega} { 1 \over e^{\beta \bar \omega}  - 1},
\end{equation}
where, obviously, we retain the same notations for the integrals (4.6) and (4.7) at $k=1$, for convenience.

It is worth noting that the NP term (5.2) describes the same massive gluonic
excitation $\bar \omega$ (3.5), but its propagation, however, suppressed by the $\alpha_s$-order.
In the free PT $\alpha_s = \Delta^2=0$ limit, the above-defined composition $N(T)$ becomes zero as well as $P^g_{NP}(T)$ itself,
as it follows from eqs.~(3.1)-(3.4). Thus the gluon pressure $P_g(T)$ (5.1), satisfies to the normalization condition of the free PT vacuum to
zero, as underlined above.  At the same time, it is truly NP one, since it vanishes in the PT $\Delta^2=0$ limit as well.
Numerically calculated the gluon pressure (5.1) is shown in fig. 1. It has a maximum at some "characteristic"
temperature $T=T_c = 266.5 \ \MeV$.
Let us now analytically investigate the low-temperature (below $T_c$) behavior of the NP gluon pressure (5.1) in more detail. It will make it possible
to explicitly show the Hagedorn-type nature of the corresponding expansion in this region. At the same time,
its high-temperature (above $T_c$) behavior suffices to briefly discuss.

\begin{figure}
\begin{center}
\includegraphics[width=10cm]{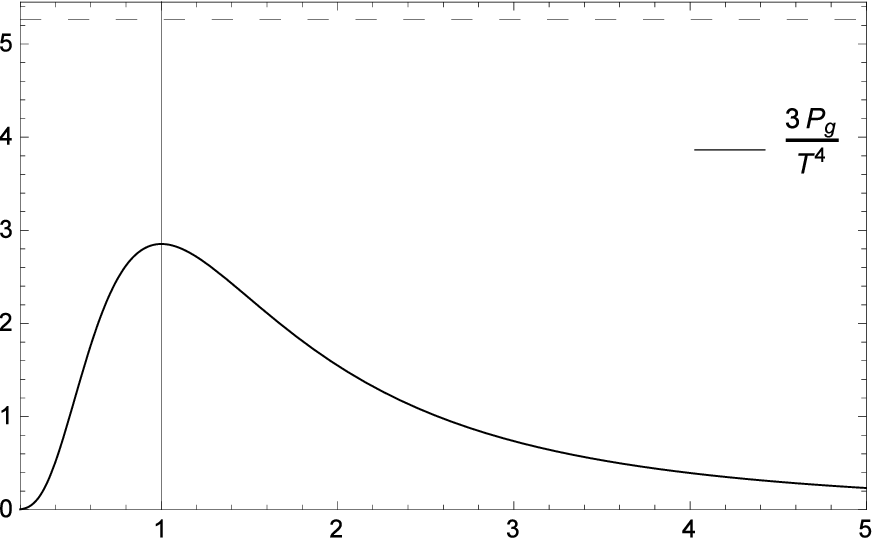}
\caption{The NP gluon pressure (5.1) scaled (i.e., divided) by $T^4 /3$ is shown as a function of $T/T_c$ (solid curve).
It has a maximum at $T=T_ c = 266.5 \ \MeV$ (vertical solid line). The horizontal dashed line is the general Stefan-Boltzmann (SB) constant
$3P_{SB}(T)/ T^4 =(24/45) \pi^2$. One can conclude that NP effects due to the mass gap are still important approximately up to $5T_c$.}
\label{fig:1}
\end{center}
\end{figure}

\section{Low-temperature expansion. The Hagedorn-type structure}

In order to investigate the behavior of the NP gluon pressure (5.1) in the low-temperature region ($T \leq T_c$) it is convenient to present it as follows:

\begin{equation}
P_g(T) = {6 \over \pi^2} \Delta^2 P_1 (T) + {16 \over \pi^2} T N(T) + P^s_{NP}(T),
\end{equation}
using eqs.~(3.1)-(3.4) and integral (5.2). Let us note
that in all these integrals the variable $e^{-\beta\omega}$ with the replacements $\omega \rightarrow \omega', \bar \omega$
is always small in this region, especially in the  ($T \rightarrow 0, \ \beta = T^{-1} \rightarrow \infty$) limit.
So one can expand the corresponding mean numbers (3.6) in the form of the corresponding Taylor series \cite{21} as follows:

\begin{equation}
 N_g \equiv N_g(\beta, \omega) = {1 \over e^{\beta\omega} -1} = e^{- \beta\omega} (1 - e^{-\beta\omega})^{-1} = \sum_{n=1}^{\infty} e^{- n \beta\omega}
\end{equation}
and

\begin{equation}
\ln \left( 1 - e^{- \beta \omega} \right) = - \sum_{n=1}^{\infty} {1 \over n} e^{- n \beta\omega}
\end{equation}
with the above-mentioned replacements (here $n$ is simply summation index and do not mix it up with the Matsubara frequencies labeled by $n$
in eq.~(2.1) and eqs.~(2.3)-(2.5)).
After substitution of these series into the corresponding integrals, such obtained terms
can be explicitly integrated termwise, since  the Taylor series (6.2) and (6.3) are convergent in this temperature
region and integrals calculated in this section are not divergent.

Let us begin with pointing out in advance that all exactly calculated integrals, discussed below
can be found in \cite{21,22}.  So the integral $P_1(T)$ defined in eq.~(3.2) becomes

\begin{equation}
P_1(T) = \int_{\omega_{eff}}^{\infty} \dd \omega \ \omega \ N_g(\beta, \omega) =
\int_{\omega_{eff}}^{\infty} \dd \omega \ \omega \sum_{n=1}^{\infty} e^{- n \beta\omega} =
\sum_{n=1}^{\infty} \int_{\omega_{eff}}^{\infty} \dd \omega \ \omega  e^{- n \beta\omega}.
\end{equation}
The almost trivial integration yields

\begin{equation}
P_1(T) = \sum_{n=1}^{\infty} \left( {1 \over n^2}T^2 + {1 \over n} \omega_{eff} T \right) e^{ - n {\omega_{eff} \over T }}.
\end{equation}

The integral $P_2(T)$ defined in eqs.~(3.4) can be considered in the same way after the substitution of the expansion (6.3), so it becomes

\begin{equation}
P_2(T) = \int_{\omega_{eff}}^{\infty} \dd \omega \ \omega^2
\ln \left( 1- e^{-\beta\omega} \right) = - \sum_{n=1}^{\infty} {1 \over n} \int_{\omega_{eff}}^{\infty} \dd \omega \ \omega^2
\ e^{- n \beta\omega},
\end{equation}
and exactly integrating it, one obtains

\begin{equation}
P_2(T) = - \sum_{n=1}^{\infty} {1 \over n}  \left( {2 \over n^3} T^3 + {2 \over n^2} \omega_{eff} T^2 + {1 \over n} \omega^2_{eff} T \right)
e^{ - n {\omega_{eff} \over T }}.
\end{equation}

The integral $P_3(T)$ defined in eqs.~(3.4) after the substitution of the expansion (6.3) with the replacement
$\omega \rightarrow \omega'$ looks like

\begin{equation}
P_3(T)= \int_0^{\omega_{eff}} \dd \omega \ \omega^2 \ln
\left( 1 - e^{- \beta\omega'} \right) = - \sum_{n=1}^{\infty} {1 \over n} \int_0^{\omega_{eff}} \dd \omega \ \omega^2
e^{- n \beta\omega'}.
\end{equation}
Replacing the variable $\omega$ by the variable $\omega'$ in accordance with the relation (3.5), this integral becomes

\begin{equation}
P_3(T) = - \sum_{n=1}^{\infty} {1 \over n} \int_a^{\omega'_{eff}} \dd \omega' \ \omega'
\sqrt{(\omega'^2 - a^2)} \ e^{- n \beta\omega'},
\end{equation}
where

\begin{equation}
\omega'_{eff} = \sqrt{(\omega_{eff}^2 + a^2)}, \ \quad a = \sqrt{3} \Delta.
\end{equation}

Noting further that the variable $x=a^2 / \omega'^2 \leq 1$, we can formally expand

\begin{equation}
\sqrt{(\omega'^2 - a^2)}= \omega'( 1 - x)^{1/2} = \omega' \left[ 1 - {1 \over 2} {a^2 \over \omega'^2} + \sum_{k=2}^{\infty}{1/2  \choose k}(-x)^k \right],
\end{equation}
then from the last integral it follows

\begin{equation}
P_3(T) = - \sum_{n=1}^{\infty} {1 \over n} \int_a^{\omega'_{eff}} \dd \omega' \ \omega'^2 \ e^{- n \beta\omega'}
+ {3 \over 2} \Delta^2  \sum_{n=1}^{\infty} {1 \over n} \int_a^{\omega'_{eff}} \dd \omega' \ e^{- n \beta\omega'}  - \sum_{n=1}^{\infty} {1 \over n}
P^{(n)}_3(T),
\end{equation}
where

\begin{equation}
P^{(n)}_3(T) = - \int_a^{\omega'_{eff}} \dd \omega' \ \omega'^2 \ e^{- n \beta\omega'}
\sum_{k=2}^{\infty}{1/2  \choose k}(-x)^k.
\end{equation}

Let us consider the last integral (6.13) in more detail. Since the series over $k$ are convergent in the interval
of integration and the functions depending on $k$ are integrable in this interval, these series may be integrated termwise \cite{18}, that is,

\begin{equation}
P^{(n)}_3(T) = - \sum_{k=2}^{\infty}{1/2  \choose k}(-a^2)^k \int_a^{\omega'_{eff}} \dd \omega'
\ { e^{- n \beta\omega'} \over (\omega')^{2k-2}}.
\end{equation}
Integrating it, one obtains

\begin{equation}
P^{(n)}_3(T) = - \sum_{k=2}^{\infty}{1/2  \choose k}(-a^2)^k \left[ N_3^{(n,k)}(T, \omega') \right]_a^{\omega'_{eff}}
\end{equation}
and $\left[ N_3^{(n,k)}(T, \omega') \right]_a^{\omega'_{eff}}$ denotes the result of the integration
over $\omega'$ in eq.~(6.14) in the interval $[a, \omega'_{eff}]$, while the function $N_3^{(n,k)}(T, \omega')$ itself is

\begin{equation}
N^{(n,k)}_3(T, \omega') = - e^{- n \beta\omega'} \sum_{m=1}^{2k-3}{ (- n \beta)^{m-1} (\omega')^{m +2 - 2k}
\over (2k-3)(2k-4)...(2k-2-m)} + { (- n \beta)^{2k-3} \over (2k-3)!} {\rm Ei}(- n \beta\omega'), \quad k=2,3,4,...
\end{equation}
The series for the exponential integral function ${\rm Ei}(- n \beta\omega')$ is \cite{18}

\begin{equation}
{\rm Ei}(- n \beta\omega') =  e^{- n \beta\omega'} \sum_{l=1}^p (-1)^l { (l-1)! \over (n \beta\omega')^l} + R_p,
\end{equation}
where the relative error in the expansion (6.17) should satisfy $\mid{R_p}\mid < p! / (n \beta\omega')^{p+1}$ for real numbers.
If one chooses $p = 2k-4$ in the previous equation and correspondingly adjusting the relative error $R_p$, it is easy to show that both terms
in eq.~(6.16) for $N^{(n,k)}_3(T, \omega')$ cancel each other termwise for any $k \geq 2$, and thus

\begin{equation}
N^{(n,k)}_3(T, \omega') = 0, \quad k=2,3,4,...
\end{equation}
which leads to

\begin{equation}
P^{(n)}_3(T) = 0
\end{equation}
via eq.~(6.15). Equivalently, we can choose $p = 2k-3$ and neglecting $R_p$, then both terms
in eq.~(6.16) for $N^{(n,k)}_3(T, \omega')$ again will cancel each other termwise for any $k \geq 2$, by neglecting the term
of the same order as $R_p$ in the first sum of eq.~(6.16). Going back to eq.~(6.12) and easily integrating the first two terms,
and taking into account the previous result, one comes to the following expression, namely

\begin{eqnarray}
P_3(T) &=& \sum_{n=1}^{\infty} {1 \over n} \left( {2 \over n^3} T^3 + {2 \over n^2} \omega'_{eff} T^2 + {1 \over n} \omega'^2_{eff} T \right)
e^{ - n {\omega'_{eff} \over T }}
- \sum_{n=1}^{\infty} {1 \over n} \left( {2 \over n^3} T^3 +  {2 \over n^2} a T^2 + {1 \over n} a^2 T \right) e^{ - n { a \over T }} \nonumber\\
&-& { 1 \over 2} a^2 T  \sum_{n=1}^{\infty} {1 \over n^2} \left(  e^{ - n {\omega'_{eff} \over T }}
-  e^{ - n { a \over T }} \right).
\end{eqnarray}

The integral $P_4(T)$ defined in eqs.~(3.4) after the substitution of the expansion (6.3)  with the replacement $\omega \rightarrow \bar \omega$
looks like

\begin{equation}
P_4(T) = \int_0^{\infty} \dd \omega \ \omega^2 \ln \left( 1 -
e^{- \beta \bar \omega} \right) = - \sum_{n=1}^{\infty} {1 \over n} \int_0^{\infty} \dd \omega \ \omega^2
e^{- n \beta \bar \omega},
\end{equation}
and replacing the variable $\omega$ by the variable $\bar \omega$ in accordance with the relation (3.5),
this integral becomes

\begin{equation}
P_4(T) = - \sum_{n=1}^{\infty} {1 \over n} \int_{(a/2)}^{\infty} \dd \bar \omega \ \bar \omega
\sqrt{(\bar \omega^2 - (a/2)^2)} \ e^{- n \beta \bar \omega}.
\end{equation}
Comparing eq.~(6.9) with this eq.~(6.22), one can conclude that the last one is the first one by putting
formally $\omega'_{eff} = \infty$ and replacing $a \rightarrow a/2$. Doing so in the expansion (6.20), for integral (6.22) one finally obtains

\begin{equation}
P_4(T) =
- \sum_{n=1}^{\infty} {1 \over n} \left( {2 \over n^3} T^3 +  {a \over n^2} T^2 + {a^2 \over 4 n} T \right) e^{ - n { a \over 2 T }}
+ { 1 \over 8} a^2 T  \sum_{n=1}^{\infty} {1 \over n^2} e^{ - n { a \over 2 T }}.
\end{equation}

Let us now consider eq.~(5.2), which after the substitution of the expansion (6.2) with the replacement $ \omega \rightarrow \bar \omega$
becomes

\begin{equation}
P^s_{NP}(T) = {9 \alpha_s \over 2 \pi^2} \Delta^2
\int_{\Lambda_{YM}}^{\infty} \dd\omega \ \omega^2 \
{ 1 \over \bar \omega} { 1 \over e^{\beta \bar \omega}  - 1} = {9 \alpha_s \over 2 \pi^2} \Delta^2
\sum_{n=1}^{\infty} \int_{\Lambda_{YM}}^{\infty} \dd\omega \ \omega^2 \
{ 1 \over \bar \omega}  e^{- n \beta \bar \omega},
\end{equation}
and $\bar \omega$ is given by the relation (3.5). Replacing the variable $\omega$ by the variable
$\bar \omega$, one obtains

\begin{equation}
P^s_{NP} (T) = {9 \alpha_s \over 2 \pi^2} \Delta^2  \sum_{n=1}^{\infty} \int_{\tilde{\omega}_{eff}}^{\infty} \dd \bar \omega \
\sqrt{(\bar \omega^2 - (a/2)^2)} \ e^{- n \beta \bar \omega}, \quad \tilde{\omega}_{eff} = \sqrt{\Lambda^2_{YM} + (a/2)^2},
\end{equation}
and for $a$ see eq.~(6.10). Noting that the variable $ z= a^2 / 4 \bar \omega^2 < 1$ in this case, we can use the expansion like (6.11),
taking into account only the substitution $a \rightarrow a/2$, in order to obtain

\begin{equation}
P^s_{NP}(T) = {9 \alpha_s \over 2 \pi^2} \Delta^2 \sum_{n=1}^{\infty} \left[ \int_{\tilde{\omega}_{eff} }^{\infty} \dd \bar \omega \ \bar \omega \
e^{- n \beta \bar \omega}
 -{ 1 \over 8} a^2 \int_{\tilde{\omega}_{eff}}^{\infty} \dd \bar \omega \ {e^{- n \beta \bar \omega} \over \bar \omega} + P_s^{(n)}(T) \right].
\end{equation}
Due to the same formalism which has been used previously in order to get the result (6.19), one can conclude
that $P_s^{(n)}(T)=0$ as well. Easily integrating the first two terms, one comes to the following expansion

\begin{equation}
P^s_{NP}(T) = {9 \alpha_s \over 2 \pi^2} \Delta^2 \sum_{n=1}^{\infty} \left[ \left( {1 \over n^2} T^2 + {1 \over n} \tilde{\omega}_{eff}T \right)
 e^{- n {\tilde{\omega}_{eff} \over T}} + {1 \over 8} a^2 {\rm Ei}( - n {\tilde{\omega}_{eff} \over T}) \right],
\end{equation}
where the corresponding exponential integral function is defined by eq.~(6.17) and

\begin{equation}
\tilde{\omega}_{eff} = \sqrt{\Lambda^2_{YM} + (3/4) \Delta^2}.
\end{equation}

Collecting all our results of the corresponding integrations and after some re-arrangement of the terms, as well as introducing
the explicit dependence on the mass gap with the help of the relations (3.7) and (6.10), for the NP gluon pressure (6.1) or, equivalently, (5.1),
one finally obtains

\begin{eqnarray}
P_g(T) &=& {6 \over \pi^2} \Delta^2 T^2 \sum_{n=1}^{\infty} {1 \over n^2} \left[ \left( 1 + 1.48 n {\Delta \over T} \right) e^{ - 1.48 n (\Delta / T )}
- 4 \left( e^{ - 2.28 n (\Delta / T )} - e^{ - \sqrt3 n (\Delta / T )} \right) - e^{ - (\sqrt3 /2) n (\Delta / T )} \right] \nonumber\\
&+& {16 \over \pi^2} T^4 \sum_{n=1}^{\infty} {1 \over n^2} \left[ \left( {2 \over n^2} + {4.56 \over n}{\Delta \over T } + 5.19 { \Delta^2 \over T^2}
\right)  e^{ - 2.28 n (\Delta / T )} - \left( {2 \over n^2} + {2.96 \over n}{ \Delta \over T } + 2.19 { \Delta^2 \over T^2}
\right)  e^{ - 1.48 n (\Delta / T )} \right] \nonumber\\
&-& {16 \over \pi^2} T^4 \sum_{n=1}^{\infty} {1 \over n^2} \left[ \left( {2 \over n^2} + {2 \sqrt3 \over n}{ \Delta \over T } + 3{ \Delta^2 \over T^2}
\right)  e^{ - \sqrt3 n (\Delta / T)} - \left( {2 \over n^2} + {\sqrt3 \over n}{ \Delta \over T } + {3 \over 4}{ \Delta^2 \over T^2}
\right)  e^{ - (\sqrt3 /2 ) n (\Delta / T )} \right] \nonumber\\
&+& {9 \over 2 \pi^2} \alpha_s \Delta^2 T^2 \sum_{n=1}^{\infty} {1 \over n^2} \left[ \left( 1
+ { n \tilde{\omega}_{eff} \over  T} \right) e^{- n (\tilde{\omega}_{eff} / T)}
+ {3 \over 8} n^2 {\Delta^2 \over T^2} {\rm Ei}( - n (\tilde{\omega}_{eff} / T)) \right], \quad T \leq T_c,
\end{eqnarray}
where $\tilde{\omega}_{eff}$ is given in eq.~(6.28).

The expression (6.29) is nothing else but the Hagedorn-type expansion for the NP gluon pressure in the low-temperature region.
Its effective gluonic excitations are mainly expressed in terms of the mass gap.
It is dynamically generated by the strong self-interaction of
massless gluon modes, and thus is responsible for all the NP effects in the YM ground-state at any temperature \cite{8}.
All these effective gluonic excitations are of the NP origin. They vanish from GM spectrum in the PT $\Delta^2=0$ limit.
Other interesting features of the expansion (6.29) are: a non-analytical dependence on
the mass gap $\Delta^2$ in some terms $\sim \Delta^3$ and $\sim \Delta$.
The PT correction of the $\alpha_s$-order depends
on the mass gap squared analytically. The presence of terms $\sim T^4$, being, nevertheless, of the NP origin, since the overall coefficient
in front of them becomes zero in the PT $\Delta^2=0$ limit (as underlined above). This is in agreement with the initial normalization condition
of the free PT vacuum to zero.

It is instructive to show this expansion as a function of the variable $T_c/T$.
It suffices to do this by introducing the corresponding number of the exponents, using the numerical values of the mass gap, characteristic
temperature and $\Lambda_{YM}$, respectively, namely $\Delta = 0.6756 \ \GeV$, $T_c = 0.2665 \ \GeV$ and $\Lambda_{YM} = 0.3 \ \GeV$,
and taking into account the relation (6.28). Such kind of the expansion looks like

\begin{eqnarray}
P_g(T) &=& {6 \over \pi^2} \Delta^2 T^2 \sum_{n=1}^{\infty} {1 \over n^2} \left[ \left( 1 + 1.48 n {\Delta \over T} \right) e^{ - 3.75 n ( T_c / T )}
- 4 \left( e^{ - 5.78 n (T_c / T )} - e^{ - 4.39 n (T_c / T )} \right) - e^{ - 2.19 n (T_c / T )} \right] \nonumber\\
&+& {16 \over \pi^2} T^4 \sum_{n=1}^{\infty} {1 \over n^2} \left[ \left( {2 \over n^2} + {4.56 \over n}{\Delta \over T } +5.19 { \Delta^2 \over T^2}
\right)  e^{ - 5.78 n (T_c / T )} - \left( {2 \over n^2} + {2.96 \over n}{ \Delta \over T } + 2.19 { \Delta^2 \over T^2}
\right)  e^{ - 3.75 n (T_c / T )} \right] \nonumber\\
&-& {16 \over \pi^2} T^4 \sum_{n=1}^{\infty} {1 \over n^2} \left[ \left( {2 \over n^2} + {2 \sqrt3 \over n}{ \Delta \over T } + 3{ \Delta^2 \over T^2}
\right)  e^{ - 4.39 n (T_c / T)} - \left( {2 \over n^2} + {\sqrt3 \over n}{ \Delta \over T } + {3 \over 4}{ \Delta^2 \over T^2}
\right)  e^{ - 2.19 n (T_c / T )} \right] \nonumber\\
&+& {9 \over 2 \pi^2} \alpha_s \Delta^2 T^2 \sum_{n=1}^{\infty} {1 \over n^2} \left[ \left( 1
+ { n \tilde{\omega}_{eff} \over  T} \right) e^{- 2.46 n (T_c / T)}
+ {3 \over 8} n^2 {\Delta^2 \over T^2} {\rm Ei}( - 2.46 n (T_c / T)) \right], \quad T \leq T_c.
\end{eqnarray}
So close to $T_c$ this expansion shows an exponential rise in the number of dynamical degrees of freedom in the $T \rightarrow T_c$ limit,
explicitly seen in fig. 1. In the opposite $T \rightarrow 0$
limit the gluon pressure is exponentially suppressed. The maximum of temperature at which the Hagedorn-type expansion
(6.30) is valid is $T_c$, then it makes sense to identify $T_c$
with the Hagedorn-type transition temperature $T_h$, i.e., to put $T_h = T_c$ within our approach (and see discussion in sect. VIII as well).
It is worth underlying once more that the Hagedorn-type expansion (6.29)
or, equivalently, (6.30) is nothing else but the NP gluon pressure (5.1) in the low-temperature region $T \leq T_h=T_c$.

Concluding this part, let us stress that the Hagedorn-type pressure (6.29),
and hence (6.30), is closely related to the asymptotic of the gluon mean number (6.2) in the low-temperature region $T \leq T_c= T_h$,
and on account of the replacements $\omega \rightarrow \omega', \bar \omega$ in it.
It is even possible to say that the Hagedorn-type structure of the expansion (6.30) is determined by them
in this temperature interval within the mass gap approach to QCD at finite temperature. In other words, it was not introduced by hand, but
it was due to the corresponding asymptotics of the gluon mean numbers and the structure of the NP gluon pressure 86.1) as a function of the mass gap.

\section{High-temperature expansion. The polynomial structure}

In order to investigate the behavior of the NP gluon pressure (5.1) in the high-temperature region ($T \geq T_c$),
it is convenient to re-write it as follows:

\begin{equation}
P_g(T) = \Delta^2 T^2 - {6 \over \pi^2} \Delta^2 P'_1 (T) + {16 \over \pi^2} T N(T)  + P^s_{NP}(T) + P^s_{PT}(T),
\end{equation}
where

\begin{equation}
P'_1(T) = \int^{\omega_{eff}}_0 \dd \omega \ \omega \ N_g(\beta, \omega) =
\int^{\omega_{eff}}_0 \dd \omega {\omega \over e^{\beta\omega} -1}.
\end{equation}
It is easy to show that the expressions (6.1) and (7.1) are the same, because of the relations
$P'_1(T) = (\pi^2 / 6) T^2 - P_1 (T), \ \int^{\infty}_0 (\dd \omega \omega / e^{\beta\omega} -1) = (\pi^2 / 6) T^2$,
where the integral $P_1 (T)$ is explicitly given in eq.~(3.2).
At moderately high temperatures up to approximately a few $T_c$ the exact functional
dependence on the mass gap $\Delta^2$ and temperature $T$ of the NP gluon pressure (7.1)
remains rather complicated. From fig. 1 it follows that the NP effects due to the mass gap are still important up to rather high temperature,
estimated as $(4-5)T_c$. The gluon pressure has a polynomial character in integer powers of $T$ up to
$T^2$ at very high temperatures only (see below). As mentioned above, it is related to the corresponding asymptotic of the gluon mean number
(3.6). In the high-temperature limit $T \rightarrow  \infty \ (\beta = T^{-1} \rightarrow 0)$, the gluon mean number
$N_g(\beta, \omega)$ can be reproduced by the corresponding series in powers of $(\beta\omega)$ if the variable $\omega$ is restricted, namely

\begin{equation}
N_g(\beta, \omega) = { 1 \over e^{\beta\omega} -1} = (\beta\omega)^{-1} [ 1 - {1 \over 2} (\beta\omega) + O(\beta^2)],
\quad \beta \rightarrow 0,
\end{equation}
with the corresponding replacements $\omega \rightarrow \omega', \bar \omega$.
It is worth noting in advance that in what follows for our purpose it is sufficient to keep only the positive
powers of $T$ in the evaluation of the high-temperature expansion for the NP gluon pressure (7.1).
Omitting all these tedious derivations, which can be explicitly found in \cite{8,10},
the high-temperature expansion for the gluon pressure (7.1) up to the leading and next-to-leading orders, is as follows:

\begin{eqnarray}
P_g(T) &\sim& {12 \over \pi^2} \Delta^2 \omega_{eff}T  + {8 \over 3 \pi^2} \omega^3_{eff} T \ln \left( { \omega'_{eff}\over \bar \omega_{eff} } \right)^2 \nonumber\\
&+& {2 \sqrt{3} \over \pi^2} \Delta^3 T\arctan \left({ 2 \omega_{eff} \over \sqrt{3} \Delta} \right) - {16 \sqrt{3} \over \pi^2} \Delta^3 T \arctan \left({ \omega_{eff} \over \sqrt{3} \Delta} \right) \nonumber\\
&+& {9 \over 2 \pi^2} \alpha_s \Delta^2 \left[ {\pi^2 \over 6} T^2 -
T \left( \Lambda_{YM} - {\sqrt{3} \over 2} \Delta \arctan \left( {2 \Lambda_{YM} \over \sqrt{3} \Delta} \right) \right) \right],
\quad T \rightarrow \infty,
\end{eqnarray}
where $\bar \omega_{eff} = \sqrt{\omega^2_{eff} + (3/4) \Delta^2}$, while $\omega_{eff}$ and $\omega'_{eff}$ are shown in (3.7)
and (6.10), respectively. Here it suffices to express the gluon pressure
in terms of the above-mentioned effective $\omega_{eff}$'s and the mass gap itself.

A non-analytical dependence on the mass gap occurs in terms $\sim \Delta^3$, though
$\Delta^2$ is not an expansion parameter like $\alpha_s$ is in hot PT QCD, where a non-analytical dependence on $\alpha_s$
has been discovered (see, for example \cite{23} and references therein). The term $\sim T^2$ has been first introduced
in the phenomenological equation of state (EoS) \cite{24} and widely discussed in \cite{7,8,25,26,27,28,29,30,31,32}.
On the contrary, in our approach both terms $\sim T^2$ and $\sim T$ have not been
introduced by hand. They naturally appear on a general ground as a result of the
explicit presence of the mass gap from the very beginning in our EoS (7.1).

It is interesting to note that the mass scale parameter in the leading NP term $\sim T^2$ in the expansion (7.4) is
$ (9 / 2 \pi^2) \times (\pi^2 / 6) \Delta^2 = ( 3 / 4) \Delta^2 = \bar m^2_{eff}$ due to the relations (3.5). Its numerical value is
$\bar m_{eff} = 585 \ \MeV$. The scale of the NP dynamics investigated in \cite{24} is $ M = 596 \ \MeV$
at almost the same $T_c$ as ours, namely $T_c = 270 \ \MeV$. 
Also our lowest effective massive excitation $\bar m_{eff}$
is in a good numerical agreement with the Debye screening mass, estimated as $\sim (500-600) \ \MeV$ \cite{12,23}. 
It may or may not be a coincidence, but these numbers
are very close to each other, though obtained by different approaches.

The appearance of the NP massive gluonic excitation $\bar m^2_{eff} =( 3 / 4) \Delta^2$
in this expansion (though suppressed as it should be, see remarks below) is clear evidence of the importance
of the NP effects up to a few $T_c$, as it has been underlined above, see fig. 1. However,
they play an important role in the whole temperature region (see sect. V). In the Hagedorn-type expansions (6.29)-(6.30)
and in eq.~(7.4) they explicitly appear by the substitutions  $\Delta \rightarrow (1 / \sqrt{3}) m'_{eff}$ and
$\Delta \rightarrow (2 / \sqrt{3}) \bar m_{eff}$, as it follows from eqs.~(3.5).
They are suppressed in different ways in the limit of very high temperatures only, as it can be concluded from the expansion (7.4).

A few important issues concerning the high-temperature asymptotic of the gluon pressure (7.4) are to be discussed in more detail.
The corresponding expansion for the composition $(16 / \pi^2) T N_1(T)= (16 / \pi^2) T [ P_2(T)-P_4(T)]$, which enters
the composition $N(T)$ in (7.1), is as follows:

\begin{eqnarray}
{16 \over \pi^2} T N_1(T) &\sim& -2P_{SB}(T) + 2 P_{SB}(T) - \Delta^2 T^2 + {6 \over \pi^2} \Delta^2 \omega_{eff} T
- {16 \over \pi^2} T P_4^{(2)}(T) \nonumber\\
&\sim& - \Delta^2 T^2 + {6 \over \pi^2} \Delta^2 \omega_{eff} T - {16 \over \pi^2} T P_4^{(2)}(T),
\quad T \rightarrow \infty,
\end{eqnarray}
where the expression for the integral $P_4^{(2)}(T)$ is not important for present discussion.
So one can conclude that at high temperatures
the exact cancelation of the $P_{SB}(T)$ terms occurs within this composition.
On the other hand, substituting it into eq.~(7.1) the cancelation of the $\Delta^2 T^2$ term occurs within the pressure $P_g(T)$ itself.
Let us emphasize once more that the SB term disappears from the gluon pressure (7.1) above $T_c$
due to the normalization of the free PT vacuum to zero from the very beginning. The cancellation of the truly NP terms $\Delta^2 T^2$
simply shows that exact $T^2$ behavior cannot start just from $T_c$ due to the rather complicated dependence of the gluon pressure
on the mass gap and temperature in the moderately high temperature interval (approximately up to $5T_c$, see fig. 1).
It would be very surprised if a pure NP contribution were survived in the limit of very high temperature,
while for its PT correction it would be expected/possible. In other words, the $\Delta^2 T^2$ behavior of $P_g(T)$ in (7.1) is replaced
by $\sim \alpha_s \Delta^2 T^2$ behavior in (7.4) only in this limit. At the same time, the second purely NP term $\sim T$ is
suppressed in comparison with the first term in the high temperature limit, indeed.

Nevertheless, the approximate $\sim T^2$ behavior (i.e., not suppressed by the $\sim \alpha_s$-order)
up to the rather high temperature of such thermodynamic quantity as the trace anomaly or, equivalently, the interaction
measure $I(T)=\epsilon(T)-3P(T)$ is likely to appear, since it depends on the derivative of the full pressure (discussed below in some details).
The interaction measure $I(T)$ is free from all the types of the purely PT contributions, by construction (and thus is very sensitive to
the truly NP effects, for preliminary discussion see \cite{33}).

Concluding, in a more compact form the previous expansion (7.4) looks like

\begin{equation}
P_g(T) = \alpha_s (3 / 4) \Delta^2 T^2 + [B_3 \Delta^3 + GeV^3]T + ...,  \quad T \rightarrow \infty,
\end{equation}
where the expressions for both constants $B_3$ and $GeV^3$ (which becomes zero in the PT $\Delta^2=0$ limit)
can be easily restored from the expansion (7.4), if necessary.

\section{Discussion and conclusions}

The NP gluon pressure (5.1) has a few interesting features. First of all, below
$T_c$ it is exponentially suppressed in the $T \rightarrow 0$ limit, see expansions (6.29) and (6.30) in this limit.
Its the most important feature is that at low temperatures  $T \leq T_c$
it is nothing else but the Hagedorn-type exponential series (6.30) for the effective gluonic
excitations, which are expressed in terms of the mass gap, generated in its turn by the strong self-interaction of massless
gluon modes. It is the only one which determines the NP dynamics in the GM within our approach \cite{8}. Nevertheless, it
plays a crucial role in the structure of the gluon pressure (5.1) in the whole temperature range, as it can be clearly
seen throughout this investigation. We call our expansion (6.30)
the Hagedorn-type since it has exponential increasing spectrum valid only up to $T_c$.
The scale of the exponential increase determines the value of the Hagedorn temperature \cite{11,34,35}. Just this happens
in the expansion (6.30) in the $T \rightarrow T_c$ limit. This means that maximum of temperature at which the Hagedorn-type structure is valid
is $T_c$, so one has to identify $T_c$ with the Hagedorn-type transition temperature $T_h$. Indeed, there is no other choice than to put
$T_h = T_c$ in the mass gap approach to QCD at non-zero temperature. In other words, the mass gap approach makes/implies
the Hagedorn-type exponential series to be necessarily arisen in hot QCD, see also discussion in \cite{36} (and references therein).

It is instructive to point out that our dynamical degrees of freedom (expressed in terms of the mass gap, as pointed out above)
are different from those which appear in the Hagedorn pressure of the glueball gas model associated
with a sum over a number of single noninteracting, relativistic particle of the corresponding masses (low-lying glueballs, i.e.,
the bound-states of the two or three gluons). It appears in the Hagedorn-mass spectra where the mass/energy density increases exponentially
as $\exp(m/T_H)$ \cite{11,36,37}. That was a main reason why we did not identify our Hagedorn-type temperature $T_h$
with the Hagedorn temperature $T_H$. However, let us note that glueball gas model alone was unable to correctly describe the
corresponding thermodynamical lattice data below $T_c$ \cite{30,37}.
Only adding the closed bosonic string contribution \cite{38}, modelling the high-lying glueballs \cite{39,40,41,42} exponential spectrum,
success has been achieved \cite{30,43}. Also the $SU(3)$ lattice entropy density has been nicely reproduced down to $0.7T_c$
by taking into account the string-type configurations of gluon fields in this joint approach (glueball gas model plus bosonic string) \cite{37}.

The NP gluon pressure (5.1) has a maximum at some characteristic temperature $T = T_c = 266.5 \ \MeV$, see fig. 1, at which its exponential rise
in the $T \rightarrow T_c$ limit is changed to fall off at $T \geq T_c$. Its fall off just after $T_c$ is not a simple
polynomial-type one, see fig. 1. It is due to its rather complicated dependence on the temperature and mass gap in the region of high
temperatures up to approximately $(4-5)T_c$. So NP effects are still important within our approach in this temperature interval, i.e.,
the gluon plasma (GP) can be considered as still remaining strongly interacting medium in this region (we call the dynamical content of the GM
above $T_c$. as GP). Only in the limit of very high temperature $T \rightarrow \infty$ it can be considered as weakly interacting medium,
and the gluon pressure has a corresponding polynomial-type character, eq.~(7.6).

Possessing these features, the NP gluon pressure (5.1) at first sight seems to have one unpleasant "defect".
From fig. 1 it clearly follows that it will never reach the general SB constant/limit at very high temperatures after $T_c$. However, that
is not a surprise, since the SB term has been canceled in the gluon pressure from the very beginning due to the normalization condition
of the free PT vacuum to zero, as discussed in some details in sect. VII. The NP gluon pressure (5.1) may change its exponential regime below
$T_c$ only in the close neighborhood of $T_c$ in order for its full counterpart (mentioned above in sect. VII) to reach the requested SB limit at high
temperatures. The SB term cannot be added to eq.~(5.1), even multiplied by the corresponding $\Theta ((T / T_c) - 1)$-function.
In this case the full pressure will get a jump at
$T=T_c$, which is not acceptable. So some other term(s), multiplied by the corresponding $\Theta ((T_c / T) - 1)$-function,
should be added as well in order to ensure a smooth transition
across $T_c$ for the full gluon pressure (which means that some fine-tuning mechanism has to be formulated for the above-mentioned purpose).
These problems make the inclusion of the SB term into the NP gluon EoS (5.1) highly
non-trivial in order to transform it into the full gluon EoS. Only after its inclusion into eq.~(5.1) in a self-consistent way,
such obtained full equation can be called the GP pressure or the GP EoS, and denoted as $P_{GP}(T)$
(we call the dynamical content of GP below $T_c$ as GM).

In this connection, one thing has to be made perfectly clear. The NP gluon pressure (5.1) will remain an important part
of the full GP pressure. It is this which will determine the low-temperature dynamical structure of the full pressure and even will play
a significant role in it rather far away from $T_c$. Let us
emphasize once more that the NP gluon pressure $P_g(T)$ (though determined in the whole temperature range), but being the NP part of the
full pressure, is not obliged and cannot reach SB limit at very high temperature.
It is the full pressure $P_{GP}(T)$ which is obliged to approach this thermodynamical limit, and
should be a continuously growing function of temperature at any point of its
domain from zero to infinity. Thus the above-discussed
unpleasant "defect" is not a real defect at all: on the contrary, the NP gluon pressure (5.1) has a correct thermodynamic
limit at very high temperatures (7.6).
The NP effects cannot indeed survive in the regime of very high temperatures, which is governed by the SB pressure $P_{SB}(T)= (8/45)\pi^2T^4$
of non-interacting massless particles (an ideal gas limit of gluons).

The NP gluon pressure (5.1) necessarily has a Hagedorn-type structure at $T \leq T_c=T_h$ and
demonstrates rather complicated dependence on the mass gap and temperature up to approximately $(4-5)T_c$.
In the limit of very high temperature $T \rightarrow \infty$ it has a polynomial behavior consistent with the SB limit.
That is why it can serve as a basic equation for its transformation into the full GP EoS.
In the forthcoming paper we will present a general formalism (the above-mentioned fine-tuning mechanism) how to transform
the gluon pressure (5.1) into the full GP EoS in a self-consistent way (i,e., not destroying the Hagedorn-type structure below $T_c$,
providing the smooth transition across $T_c$ and approaching to the SB limit above $T_c$ from below). For this we will need the lattice data
\cite{27,30,43} for the pressure on either side of $T_c$ but close to it only.
Some preliminary attempts in this direction have been already done in \cite{8,44}.
Completing this program in much more satisfactory way, we will be able to analytically describe
YM $SU(3)$ lattice thermodynamics \cite{27,30,37,43,45,46,47,48,49}, and thus to compare it with other
analytical approaches and models.

Concluding, let us note that
to discuss the interpretation of the relevant degrees of freedom in the GP below and above  $T_c$ as well as its full dynamical content
within the mass gap approach to QCD/YM at non-zero temperature is more appropriate in the framework of the full gluon pressure.
The non-trivial gluonic field configurations of the purely PT origin due to asymptotic freedom should be also taken into account.
A few messages we would like to emphasize and convey as well are:

\hspace{1mm}

a). The Hagedorn-type structure of the pressure is of crucial importance to correctly understand and describe

\hspace{5mm} the GM dynamical content at low temperatures within any approach or model.

b). It necessary arises within the mass gap approach to QCD at finite temperature.

c). It is valid up to $T_c$ only, which implies to identify it with the Hagedorn-type temperature,

\hspace{5mm} i.e., to put  $T_c=T_h$ within our approach.

d). All the dynamical degrees of freedom  can be expressed in terms of the mass gap alone in this picture.

\begin{acknowledgments}
We thank R. Pisarski for bringing our attention to the paper \cite{24}. Our thanks also go to T. Bir\'o, T. Csorg\"o, P.
V\'an, G. Barnaf\"oldi, A. Lukacs and J. Nyiri for useful discussions, remarks and help. V.G. and A.S. are grateful to
N. Partsvania for constant help. We acknowledge support by the Hungarian National Fund (OTKA) 77816 and 31520 (P. L\'evai).
Partial support comes from "NewCompStar", COST Action MP1304. M.V. was also supported by the J\'{a}nos Bolyai Research Scholarship of the Hungarian
Academy of Sciences.
\end{acknowledgments}

\end{document}